\documentclass{ws-ijmpa}
\usepackage[super,compress]{cite}
\usepackage{graphicx}
\usepackage{here}

\def\beq{\begin{equation}}
\def\eeq{\end{equation}}
\def\bea{\begin{eqnarray}}
\def\eea{\end{eqnarray}}
\def\bq{\begin{quote}}
\def\eq{\end{quote}}

\def\nnb{\nonumber}

\def\nnb{\nonumber}
\def\la{\langle}
\def\ra{\rangle}
\def\nin{\noindent}
\def\ba{\vspace*{-0.2cm}\begin{array}}
\def\ea{\end{array}\vspace*{-0.2cm}}

\def\als{\alpha_s}

\def\gg2{ \la\alpha_s G^2 \ra}
\def\gg3{g^3f_{abc}\la G^aG^bG^c \ra}
\def\ggg4{\la\als^2G^4\ra}

\begin{document}
\markboth{Stephan Narison}{  }
%\begin{frontmatter}

%%
%%%%%%%%%%%%%%%%%%%%%%%%%%%%%%%%%%%%%%%%%%%%%%%%%
%\begin{document}
\title{$\alpha_s(\mu)$ from $M_{\chi_{0c(0b)}}-M_{\eta_{c(b)}}$@N2LO} 
 \author{Stephan Narison }
   \address{Laboratoire
Univers et Particules , CNRS-IN2P3,  \\
Case 070, Place Eug\`ene
Bataillon, 34095 - Montpellier Cedex 05, France. \\
Email address: snarison@gmail.com}
\maketitle
\pagestyle{myheadings}
\markright{QCD parameters correlations...}
\begin{abstract}
\noindent
This note complements and clarifies the results obtained in the original paper {\it QCD Parameters Correlations from  Heavy Quarkonia} \,\cite{SN18}
where, here, we present a more detailed discussion of the $\alpha_s-$results obtained @ N2LO at two different subtraction scales $\mu$=2.85 and 9.50 GeV from the (pseudo)scalar heavy quarkonia mass-spliitings $M_{\chi_{0c(0b)}}-M_{\eta_{c(b)}}$. We obtain from the $M_{\chi_{0c}}-M_{\eta_{c}}$ sum rule: 
 \bea&&\hspace*{-1cm} 
 \alpha_s(2.85)=0.262(9) \leadsto\alpha_s(M_\tau)=0.318(15)~
 \leadsto\alpha_s(M_Z)=0.1183(19)(3)\nnb
 %\label{eq:muc}
\eea
and from the $M_{\chi_{0b}}-M_{\eta_{b}}$ one: 
\bea &&\hspace*{-1cm} 
 \alpha_s(9.50)=0.180(8) \leadsto\alpha_s(M_\tau)=0.312(27)
\leadsto\alpha_s(M_Z)=0.1175(32)(3),
 \eea
in complete agreement with the world average: $\alpha_s(M_Z)=0.1181(11).$
%This complementary discussion is useful for a much better understanding of the
%results in Ref.\cite{SN18}.

%% keywords
\ccode{Pac numbers: 11.55.Hx, 12.38.Lg, 13.20-Gd, 14.65.Dw, 14.65.Fy, 14.70.Dj}  
\end{abstract}
\vspace*{0.25cm}
\nin
  \\
%\end{document}
%%%%%%%%%%%%%%%%%%%%%%%%%%%%%%%%%%
%\vspace*{-1.5cm}
 %%%%%%%%%%%%%%%%%%%%%%%%%%%%%%%%%%%
% \section*{$\alpha_s$ FROM $M_{\chi_{0c(0b)}}-M_{\eta_{c(b)}}$@N2LO}
 %%%%%%%%%%%%%%%%%%%%%%%%%%%%%%%%%%%
\section{Optimized subtraction scales}
%%%%%%%%%%%%%%%%%
%\vspace*{-0.25cm}
% \nin
 Besides the usual sum rules optimization procedure (sum rule variables and QCD continuum threshold) studied in details in Ref.\,\cite{SN18}, we deduce  from Figs. 4 and 8 of Ref.\,\cite{SN18} that the ratios of charmonium and bottomium moments are optimized respectively at the values of the subtraction scales:
 \beq
 \mu_c=(2.8\sim 2.9)~ {\rm GeV ~~~ and~~~}  \mu_b=(9\sim 10)~{\rm GeV}.
 \eeq
%%%%%%%%%%%%%%%%%%%%%%%%%%%%%%%%%
\section{$\alpha_s$ and $\la \alpha_s G^2\ra$ correlation}
%%%%%%%%%%%%%%%%%%%%%%%%%%%%%%%%%
 %%%%%%%%%%%%%%%%%%%%%%%%%%%%%%%%%%%%%%%
\begin{figure}[hbt]
\vspace*{-0.5cm}
\begin{center}
\includegraphics[width=12.cm]{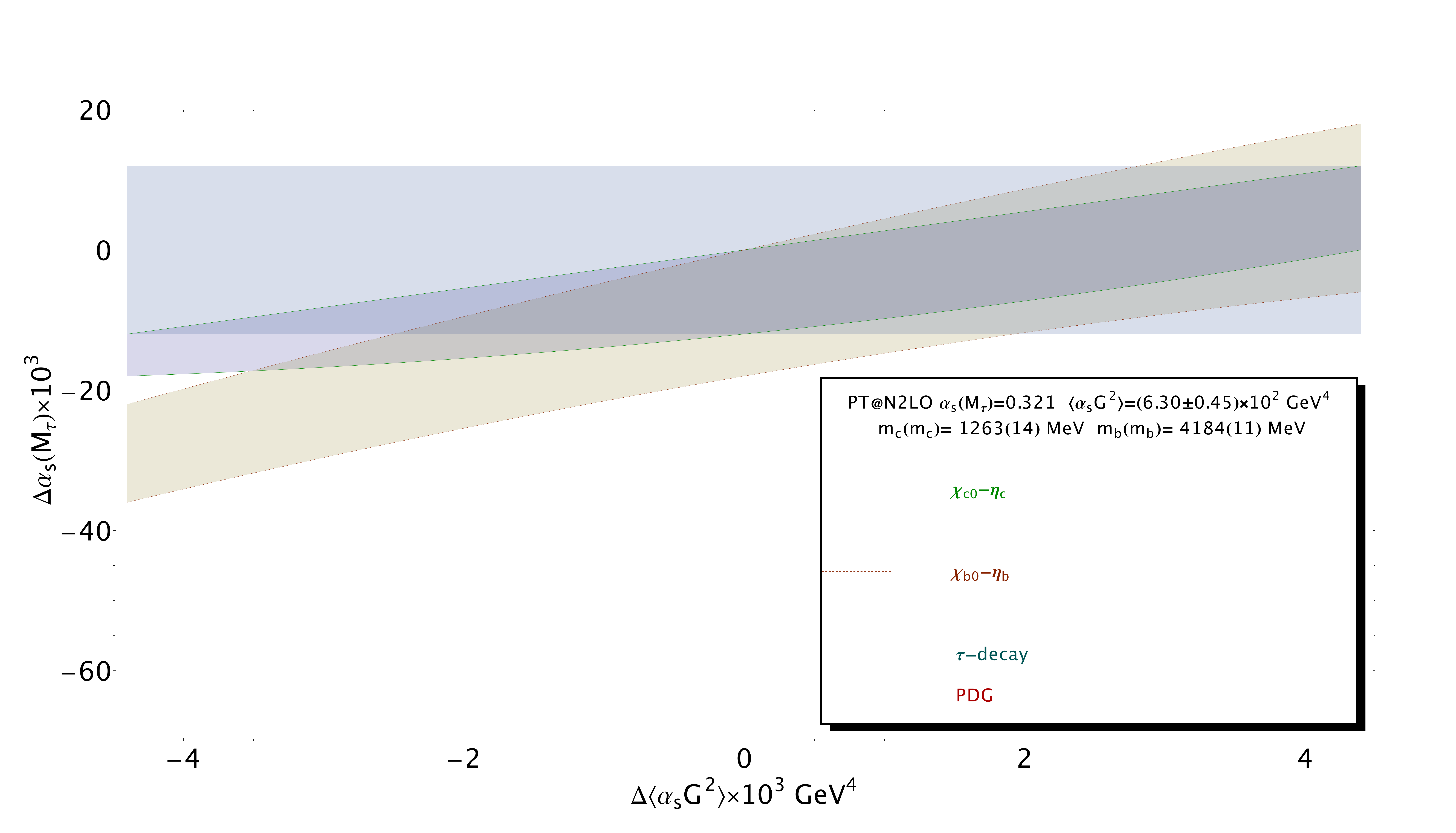}
\vspace*{-0.25cm}
\caption{\footnotesize  Correlation between $\alpha_s$ and $\la \alpha_s G^2\ra$ by requiring that the sum rules reproduce the (pseudo)scalar mass-splittings within $(2\sim 3)$ MeV.} 
\label{fig:alfas-g2}
\end{center}
%\vspace*{-0.75cm}
\end{figure} 
%%%%%%%%%%%%%%%%%%%%%%%%%%%%%%%%%%%%%%%%% 
 We study, in Fig. 15 of Ref.\,\cite{SN18} (see Fig.\ref{fig:alfas-g2}), the correlation between $\alpha_s$ and $\la \alpha_s G^2\ra$ where the charmonium (resp. bottomium) sum rules have been evaluated at $\mu_c$ (resp. $\mu_b$) but runned to the scale $M_\tau$ for a global
 comparison of the results. 
 For the range of $\la \alpha_s G^2\ra$ values allowed by different analysis ($x$-axis) and requiring that the sum rule reproduces the experimental mass-splittings  $M_{\chi_{0c}}-M_{\eta_{c}}$ by about $(2\sim 3)$ MeV, one obtains the grey band limited by the two green (continuous) curves in Fig.\ref{fig:alfas-g2} which lead to :
 \bea
&&\hspace*{-1cm} \alpha_s(2.85)=0.262(9) \leadsto\alpha_s(M_\tau)=0.318(15)~
 \leadsto\alpha_s(M_Z)=0.1183(19)(3) ~. 
 \label{eq:muc}
\eea
In the same way, the $M_{\chi_{0b}}-M_{\eta_{b}}$ bottomium sum rule evaluated at the optimization scale $\mu_b=$9 GeV gives (sand colour band limited by two dotted red curves):
\bea
 &&\hspace*{-1cm} \alpha_s(9.50)=0.180(8) \leadsto\alpha_s(M_\tau)=0.312(27)
\leadsto\alpha_s(M_Z)=0.1175(32)(3) ~.
 \eea
  %%%%%%%%%%%%%%%%%%%%%%%%%%%%%%%%%%%%%%%
\begin{figure}[hbt]
%\vspace*{-0.5cm}
\begin{center}
\includegraphics[width=12.cm]{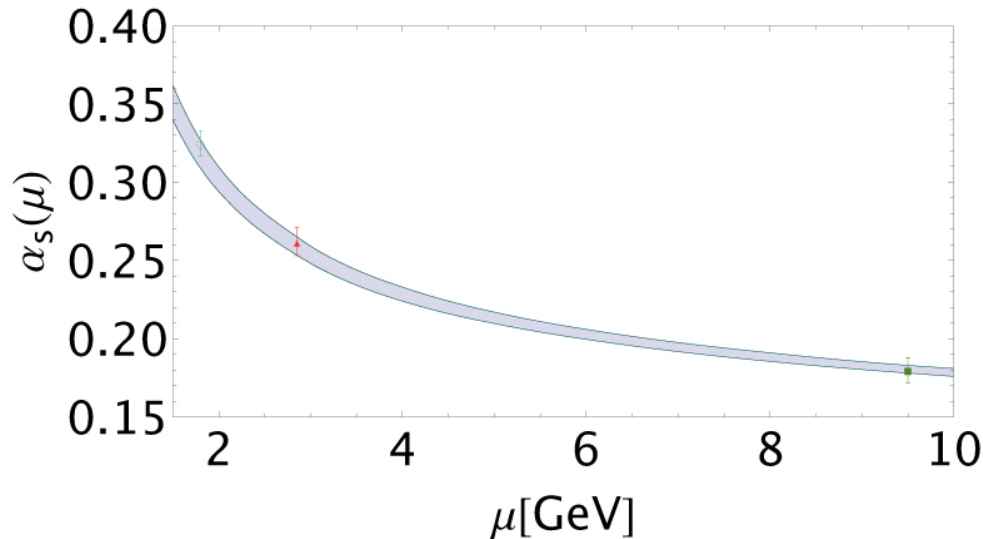}
\vspace*{-0.25cm}
\caption{\footnotesize  Comparison with the running of the world average $\alpha_s(M_Z)=0.1181(11)$\,\cite{BETHKEa,PDG} (grey band limited by the two green curves) of our predictions at three different scales: $\mu_\tau=M_\tau$  for the original $\tau$-decay width\,\cite{SNTAU} (open circle),  $\mu_c$=2.85 GeV for $M_{\chi_{c0}}-M_{\eta_c}$ (full triangle) and $\mu_b=$9.5 GeV for $M_{\chi_{b0}}-M_{\eta_b}$ (full square)\,\cite{SN18}.}
\label{fig:alfas}
\end{center}
%\vspace*{-0.75cm}
\end{figure} 
\nin
%%%%%%%%%%%%%%%%%%%%%
 \section{Comparison with the world average}
 %%%%%%%%%%%%%%%%%%%%%%%
%%%%%%%%%%%%%%%%%%%%%%%%%%%%%%%%%%%%%%%%% 
 These values of $\alpha_s(\mu)$ estimated at different $\mu$-scales are shown in Fig.\,\ref{fig:alfas} where they are compared with the running of the world average  
$\alpha_s(M_Z)=0.1181(11)$\,\cite{BETHKEa,PDG}. We have added, in the figure, your previous estimate of $\alpha_s(M_\tau)$\,\cite{SNTAU} obtained from the original $\tau$-decay rate (lowest moment)\cite{BNPa,BNPb}:
\beq
 \alpha_s(M_\tau)=0.325(8)~,
 \eeq 
 where one should note that non-perturbative corrections beyond the standard OPE (tachyonic gluon mass and duality violations)  do not affect sensibly the above value of $ \alpha_s(M_\tau)$ as indicated by the  co\"\i ncidence of the central value  with the recent one from high-moments\,\cite{PICHTAUb}.
 
 Our most precise prediction for $\alpha_s$ from the heavy-quarkonia mass-splittings comes from the (pseudo)scalar charmonium one in Eq.\,\ref{eq:muc} which corresponds to:
 \beq
 \alpha_s(M_Z)=0.1183(19)(3) ~. 
 \eeq
% \newpage
 %%%%%%%%%%%%%%%

 %%%%%%%%%%%%%%%%%%%%%%%%%%%%%%%%%%%
 \end{document}